\documentclass[12pt]{article}
\usepackage{amssymb}
\usepackage{amsmath}
\usepackage[space]{cite}
\usepackage{algorithm}
\usepackage[noend]{algpseudocode}
\usepackage[normalem]{ulem}
\usepackage{url}
\usepackage{graphicx}
\usepackage{adjustbox}
\usepackage{caption}
\usepackage{subcaption}
\usepackage{enumitem}   

\numberwithin{equation}{section}

\newcommand{\re}{{\rm e}}
\newcommand{\be}{\begin{eqnarray}}
\newcommand{\ee}{\end{eqnarray}}
\newcommand{\non}{\nonumber}
\newcommand{\id}{\mathbb{I}}

\newcommand{\myone}{\mathbf{1}}

\usepackage[usenames,dvipsnames]{color}

\newcommand{\cut}[1]{\ifmmode\text{\textcolor{red}{\sout{\ensuremath{#1}}}}\else\textcolor{red}{\sout{#1}}\fi}

\newcommand{\RN}[1]{%
  \textup{\uppercase\expandafter{\romannumeral#1}}%
}

\usepackage{quantikz}
\usetikzlibrary{circuits.logic.US}
\tikzset{
    gateO/.style={
        draw,
        circle,
        minimum width=0.5em,
        inner sep=2pt    }
}
\DeclareExpandableDocumentCommand{\gateO}{O{}{m}}{|[gateO,#1]| {#2} \qw}

\tikzset{
    gateOS/.style={
        draw,
        circle,
        minimum width=0.5em,
        inner sep=2pt,
		fill=red!20}
}
\DeclareExpandableDocumentCommand{\gateOS}{O{}{m}}{|[gateOS,#1]| {#2} \qw}

\begin{document}

\begin{titlepage}
\strut\hfill UMTG--315
\vspace{.5in}
\begin{center}

{\LARGE Qudit Dicke state preparation}\\
\vspace{1in}
\large Rafael I. Nepomechie\footnote{\tt  
nepomechie@miami.edu} and David Raveh\footnote{\tt dxr921@miami.edu}\\[0.2in] 
\large Physics Department, P.O. Box 248046, University of Miami\\[0.2in]  
\large Coral Gables, FL 33124 USA\\
\end{center}

\vspace{.5in}

\begin{abstract}
Qudit Dicke states are higher-dimensional analogues of an important
class of highly-entangled completely symmetric quantum states known as (qubit) Dicke
states.  A circuit for preparing arbitrary qudit Dicke states
deterministically is formulated.  An explicit
decomposition of the circuit in terms of elementary gates is presented, and is implemented in cirq for the qubit and qutrit cases.
\end{abstract}

\end{titlepage}

\setcounter{footnote}{0}

\section{Introduction}\label{sec:intro}

The (qubit) Dicke state $|D^{n}_{k}\rangle$ is an equal-weight 
superposition of all $n$-qubit states with $k$ $|1\rangle$'s and 
$n-k$ $|0\rangle$'s.  For example, 
\begin{equation}
|D^{4}_{2}\rangle = \frac{1}{\sqrt{6}}\left(|1100\rangle +	
|1010\rangle +	|1001\rangle +	
|0110\rangle +	|0101\rangle +	|0011\rangle\right) \,, \non
\end{equation}
where the tensor product is understood, e.g. 
$|1100\rangle = |1\rangle \otimes |1\rangle \otimes |0\rangle \otimes 
|0\rangle$. These highly-entangled states have long been studied and exploited in quantum 
information and computation for such diverse tasks as quantum networking, quantum metrology, quantum tomography,
quantum compression, and optimization, see e.g. \cite{Dicke:1954zz, Murao:1999, Childs:2000, Popkov:2004, Latorre:2004qn, 
Ozdemir:2007, 
Prevedel:2009, Toth:2010, Toth:2012, Lamata:2013, Farhi:2014, Ouyang:2014, 
Moreno:2018,
Ouyang:2021}. These states have been experimentally realized \cite{Kiesel:2007, Prevedel:2009, Wieczorek:2009}, and efficient quantum circuits for their preparation have been found \cite{Bartschi2019, Mukherjee:2020, Aktar:2021, Bartschi:2022}.
Such quantum circuits have recently been used as the starting point for 
preparing exact eigenstates of the Heisenberg spin chain \cite{VanDyke:2021kvq, VanDyke:2021nuz, 
Li:2022czv} via coordinate Bethe ansatz \cite{Bethe:1931hc, Gaudin:1983}.\footnote{Alternative 
approaches for preparing such eigenstates \cite{Nepomechie:2021bethe, Sopena:2022ntq}
via algebraic Bethe ansatz \cite{Faddeev:1996iy} do not make use of Dicke 
states.} 

There has been increasing interest in using \emph{qudits} for quantum 
computing, see e.g. the recent review \cite{Wang:2020} as well as 
\cite{Di:2011, Di:2015, Goss:2022, Hrmo:2023, Morvan:2021, Ringbauer:2022, Roy:2022} and references 
therein. Higher-dimensional analogues of qubit Dicke states, namely {\em qudit 
Dicke states} (also called generalized Dicke states, or symmetric 
basis states), have also received attention over many years, see e.g. 
\cite{Wei:2003, Popkov:2005, Hayashi:2008,  Wei:2008, Zhu:2010, Carrasco:2015sxh, Li:2021}. Let us 
consider $d$-dimensional qudits, with computational basis vectors
$|0\rangle, |1\rangle, \ldots , |d-1\rangle$ that span a 
$d$-dimensional complex vector 
space $V$. In order to specify an $n$-qudit Dicke state, it is 
convenient to introduce the notion of a \emph{multiset} 
\cite{Stanley2011}, namely, a set with repeated elements, e.g. 
$\{0,0,1,2\}$ whose element 0 has multiplicity 2. 
In particular, we define the multiset $M(\vec k)$ by
\begin{equation}
M(\vec k)	=\{ \underbrace{0, \ldots, 0}_{k_{0}}, \underbrace{1, 
\ldots, 1}_{k_{1}}, \ldots, \underbrace{d-1, \ldots, d-1}_{k_{d-1}}\} 
\,,
\label{multiset}
\end{equation}
where $k_{j}$ is the multiplicity of $j$ in $M(\vec k)$, such that $M(\vec 
k)$ has cardinality $n$. Hence, $\vec k$ is a $d$-dimensional vector such that
\begin{equation}
\vec k = (k_{0}, k_{1}, \ldots, k_{d-1})\quad \text{with}\quad k_{j} \in \{0, 
1, \ldots, n\}\quad \text{and}\quad \sum_{j=0}^{d-1} k_{j} = n\,.
\end{equation}
The corresponding $n$-qudit Dicke state is defined (see  
\cite{Wei:2003, Popkov:2005, Hayashi:2008, Wei:2008, Zhu:2010, Carrasco:2015sxh, Li:2021}) by
\begin{equation}
|D^{n}(\vec k)\rangle  = \frac{1}{\sqrt{{n \choose \vec k}}}
\sum_{w \in \mathfrak{S}_{M(\vec k)}}  | w \rangle \,,
\label{Dickedef}
\end{equation}
where $\mathfrak{S}_{M(\vec k)}$ is the set of permutations of the 
multiset $M(\vec k)$ \eqref{multiset},  
and $| w  \rangle$ is the $n$-qudit state corresponding 
to the permutation $w$; 
for example, the $n$-qudit state corresponding to the {\it identity permutation} is 
\begin{equation}
|\re(\vec k) \rangle	= |\underbrace{0 \ldots 0}_{k_{0}} \underbrace{1 
\ldots 1}_{k_{1}} \ldots \underbrace{(d-1) \ldots (d-1)}_{k_{d-1}} \rangle
= |0\rangle^{\otimes k_{0}}|1\rangle^{\otimes k_{1}}\ldots |d-1\rangle^{\otimes k_{d-1}}
\,.
\label{initialstate}
\end{equation}
Moreover, ${n \choose \vec k}$ denotes the multinomial
\begin{equation}
{n \choose \vec k} = {n \choose k_{0}, k_{1}, \ldots, k_{d-1}}	= 
\frac{n!}{\prod_{j=0}^{d-1}k_{j}!} \,,
\label{size}
\end{equation}
which is the cardinality of $\mathfrak{S}_{M(\vec k)}$.
An example with qutrits ($d=3$) is
\begin{align}
|D^{4}(2,1,1)\rangle  &= \frac{1}{\sqrt{12}}
\Big(|0012\rangle + |1002\rangle + |0102\rangle + 
|0021\rangle + |0201\rangle + |2001\rangle \non \\
&+ |0210\rangle + |0120\rangle + |1020\rangle + 
|1200\rangle + |2010\rangle + |2100\rangle \Big)\,. 
\label{d3example}
\end{align}
For the special case of qubits ($d=2$), by setting 
$\vec k = (k_{0}, k_{1}) = (n-k, k)$, we see that $|D^{n}(\vec 
k)\rangle$ reduces to the familiar Dicke state $|D^{n}_{k}\rangle$.

While properties of qudit Dicke states have been investigated 
\cite{Wei:2003, Popkov:2005, Hayashi:2008, Wei:2008, Zhu:2010, Carrasco:2015sxh, Li:2021},
the preparation of such states has not heretofore been considered.
The main goal of this paper is to formulate a circuit for 
preparing arbitrary qudit Dicke states deterministically.
Such a quantum circuit could be useful for generalizing the many 
applications of (qubit) Dicke states to qudits, such as quantum 
networking \cite{Prevedel:2009}, quantum metrology \cite{Toth:2012}, 
quantum compression \cite{Bartschi2019}, and optimization 
\cite{Farhi:2014}. In particular,
it will be needed in order to extend the algorithm 
\cite{VanDyke:2021kvq} for the (rank-1) Heisenberg spin chain
to higher-rank ($SU(d)$) integrable spin chains \cite{Sutherland:1975vr,Sutherland:1985}.

The outline of the remainder of this paper is as follows. In Sec. 
\ref{sec:Uop}, taking an approach similar to B\"artschi and Eidenbenz 
\cite{Bartschi2019} for the qubit case, we introduce a qudit Dicke 
operator $U_n$ that generates an arbitrary qudit Dicke state 
\eqref{Dickedef} from the simple 
initial state \eqref{initialstate}, and we obtain an expression 
\eqref{Uresult} for this operator as a product of certain 
$W$ operators \eqref{Wop}. The problem therefore reduces to constructing these $W$ operators in terms of elementary gates. The simplest case of qubits is considered in Sec. \ref{sec:d2},
followed by the case of qutrits in Sec. \ref{sec:d3}. Code in cirq \cite{cirq}
for simulating these circuits is included in the Supplementary Material.
The generalization to general values of $d$ is considered in Sec. 
\ref{sec:Wopgen}.
These results are briefly discussed in Sec. \ref{sec:discuss}. Matrix 
representations of the required gates and notational details
are presented in Appendix \ref{sec:matrices}.


\section{Generalities}\label{sec:Uop} 

In order to formulate a circuit for 
preparing arbitrary qudit Dicke states deterministically,
similarly to \cite{Bartschi2019} for the case $d=2$,
we begin by looking for a unitary operator $U_{n}$ (independent of $\vec{k}$) acting on $V^{\otimes n}$, 
which we call the {\em qudit Dicke operator}, that generates an arbitrary $n$-qudit Dicke state $|D^n(\vec k)\rangle$ 
\eqref{Dickedef} by acting on the identity permutation $|\re(\vec{k})\rangle$ \eqref{initialstate}
\begin{equation}
    U_n\, |\re(\vec{k})\rangle=|D^n(\vec k)\rangle\,
    \label{DickeOp}
\end{equation}
for all $\vec k$. We observe that
the qudit Dicke state \eqref{Dickedef} satisfies a recursion relation 
\begin{equation}
|D^{n}(\vec k)\rangle  = 
\sum_{s=0}^{d-1}\sqrt{\frac{k_{s}}{n}}\, |D^{n-1}(\vec k - \hat s)\rangle \otimes|s\rangle  \,,
\label{staterecursion}
\end{equation}
where $\hat s$ is a $d$-dimensional unit vector that has components $(\hat s)_j =\delta_{s,j}$, with $s=0, 1, \dots, d-1$. This recursion relation is a straightforward generalization of the $d=2$ result noted in \cite{Lamata:2013, Moreno:2018, Bartschi2019}. Let us define a corresponding operator $W_n$ (independent of $\vec{k}$) that performs the mapping
\begin{equation}
W_n |\re(\vec k)\rangle  = 
\sum_{s=0}^{d-1}\sqrt{\frac{k_{s}}{n}}\, |\re(\vec k - \hat s)\rangle \otimes|s\rangle  
\label{Wop}
\end{equation}
for all $\vec k$. Substituting \eqref{DickeOp} into both sides of  \eqref{staterecursion} and then using \eqref{Wop},
we see that the qudit Dicke operator satisfies a simple recursion in terms of the $W_n$ operator
\begin{equation}
    U_n = \left(U_{n-1} \otimes \id \right) W_n \,.
    \label{Urecursion}
\end{equation}
Using the initial condition $U_1=\id$, we can telescope the recursion \eqref{Urecursion} into a product of $W_m$ operators 
\begin{equation}
U_n =  \overset{\curvearrowright}{\prod_{m=2}^{n}} 
\left(W_m \otimes  \id^{\otimes(n-m)} \right)\,,
\label{Uresult}
\end{equation}
where the product goes from left to right with increasing $m$.
The problem of constructing qudit Dicke operators $U_n$ for any value of $d$ therefore reduces to constructing quantum circuits for the corresponding $W_{m}$ operators.

\section{The case $d=2$}\label{sec:d2} 

We begin by considering the simplest case $d=2$ (qubits), which we treat somewhat differently than \cite{Bartschi2019}. We set 
$\vec{k}=(k_0,k_1)=(n-l,l)$, so that \eqref{Wop} with $n=m$ reduces to 
\begin{equation}
W_m\, |0\rangle^{\otimes (m-l)}
|1\rangle^{\otimes l}
= 
\sqrt{\frac{m-l}{m}}\, 
|0\rangle^{\otimes (m-l-1)}
|1\rangle^{\otimes l}\otimes|0\rangle
+\sqrt{\frac{l}{m}}\, 
|0\rangle^{\otimes 
(m-l)}|1\rangle^{\otimes (l-1)}\otimes
|1\rangle  \,,
\label{Wop2}
\end{equation}

\begin{figure}[htb]
	\centering
	\begin{subfigure}{0.5\textwidth}
      \centering
\begin{adjustbox}{width=0.5\textwidth, raise=7em}
\begin{quantikz}
\lstick{$0$} & \ctrl{1}  &  \gate{R(\theta)} \vqw{1} & 
\ctrl{1}  & \qw \\
\lstick{$l=1$} & \targ{} &  \ctrl{-1} & 
\targ{}   & \qw \\
\vdots \\
\lstick{$m-1$}&\qw&\qw&\qw&\qw
\end{quantikz}
\end{adjustbox}
\caption{$\RN{1}_{m,l}$ with $l=1$}
\label{fig:I1}
    \end{subfigure}%
    \begin{subfigure}{0.5\textwidth}
        \centering
\begin{adjustbox}{width=0.5\textwidth}
\begin{quantikz}
\lstick{$0$} & \ctrl{1}  &  \gate{R(\theta)} \vqw{1} & \ctrl{1}  & \qw \\
\lstick{$1$} & \qw \vqw{2} &  \qw \vqw{2} & \qw  \vqw{2} & \qw \\
\vdots \\
\lstick{$l-1$} & \qw  & \ctrl{0} \qw & \qw  & \qw \\
\lstick{$l$} & \targ{} \vqw{-1} &  \ctrl{-1}  & \targ{} \vqw{-1}   & \qw \\
\vdots\\
\lstick{$m-1$}&\qw&\qw&\qw&\qw
\end{quantikz}
\end{adjustbox}
\caption{$\RN{1}_{m,l}$ with $l>1$}
\label{fig:Il}
	 \end{subfigure}	
\caption{Circuit diagrams for $\RN{1}_{m,l}$, with $R(\theta)$ defined in \eqref{ugate}}
\label{fig:Iops}
\end{figure}
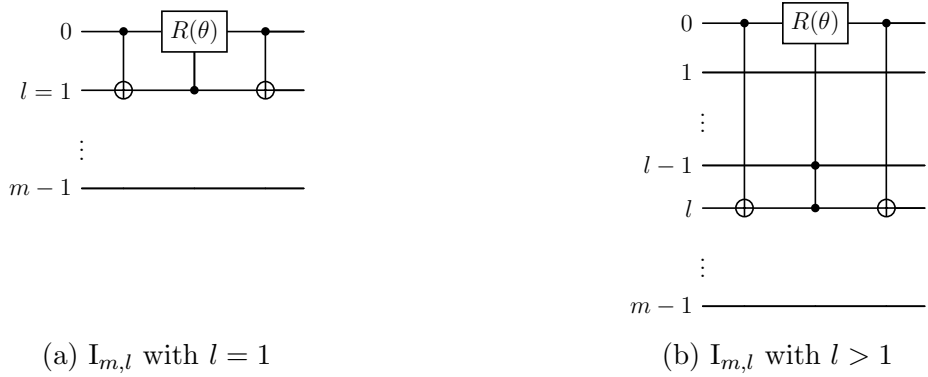

We introduce the operator $\RN{1}_{m,l}$ acting on the $l$th, $(l-1)$th, and $0$th qubit, that performs the transformation
\begin{equation}
\RN{1}_{m,l}:\quad |0\rangle_l\, |1\rangle_{l-1}\, 
|1\rangle_0 \mapsto
\sqrt{\frac{m-l}{m}}\, 
|1\rangle_l\, |1\rangle_{l-1}\, 
|0\rangle_0 +
\sqrt{\frac{l}{m}}\, 
|0\rangle_l\, 
|1\rangle_{l-1}\, 
|1\rangle_0\,,
\label{Iaction}
\end{equation}
and otherwise acts as identity (as long as the $0$th qubit is in the state $|1\rangle$, which is always the case for the input states in \eqref{Wop2}). For $l=1$, the middle qubits in \eqref{Iaction} are omitted.
The corresponding circuit diagram\footnote{The circuit 
diagrams in this paper were generated using quantikz \cite{Kay:2018}.} 
is given by Fig. \ref{fig:Iops}, with one-qubit $R^y$-gates
\begin{equation}
 R(\theta)  
 = \begin{pmatrix}
       \cos(\frac{\theta}{2}) & -\sin(\frac{\theta}{2}) \\[0.1 cm]
       \sin(\frac{\theta}{2}) &  \cos(\frac{\theta}{2})
\end{pmatrix} \,, \qquad \theta = -2 \arccos\left(\sqrt{\frac{l}{m}} \right) \,.
\label{ugate}
\end{equation}
We label $m$-qubit vector spaces from $0$ to $m-1$, going from right to left; and in circuit diagrams, the $m$ vector spaces are represented by corresponding wires labeled from the top $(0)$ to the bottom $(m-1)$,
see Appendix \ref{sec:matrices} for more details.

We note that these operators satisfy
\begin{align}
    I_{m,l'}\,|\re(m-l, l)\rangle &= |\re(m-l, l)\rangle  & \text{ if } \qquad 
    l' \ne l \,, \non \\
    I_{m,l'}\, \left[ I_{m,l}\,|\re(m-l, l)\rangle \right] &= I_{m,l}\,|\re(m-l, l)\rangle 
   & \text{ if } \qquad l' > l \,.
   \label{Iprops}
\end{align}
Hence, a quantum circuit that performs the transformation \eqref{Wop2} for all $l=1,2,\dots,m-1$
is given by an ordered product of such operators
\begin{equation}
W_m = 
\overset{\curvearrowleft}{\prod_{l=1}^{m-1}}
\RN{1}_{m,l} 	\,, 
\label{W2explicit}
\end{equation}
where the product goes from right-to-left with increasing $l$. 

The size and depth of the qubit circuit $U_n$ is ${\cal O}(n^2)$, see \eqref{size} below.

As an example with $n=6$, we see from \eqref{Uresult} and \eqref{W2explicit} that
\begin{align}
U_6 &= \left(W_2 \otimes \id^{\otimes 4}\right)
\left(W_3 \otimes \id^{\otimes 3}\right)
\left(W_4 \otimes \id^{\otimes 2} \right) 
\left(W_5 \otimes \id \right) W_6 \,, \label{U6example}\\
&= \left(\RN{1}_{2,1} \otimes \id^{\otimes 4}\right)
\left[\left(\RN{1}_{3,2} \RN{1}_{3,1} \right) \otimes \id^{\otimes 3}\right]
\left[\left(\RN{1}_{4,3} \RN{1}_{4,2} \RN{1}_{4,1} \right) \otimes \id^{\otimes 2}\right]
\left[\left( \textcolor{red}{\RN{1}_{5,4}} \RN{1}_{5,3} \RN{1}_{5,2} \textcolor{red}{\RN{1}_{5,1}} \right) \otimes \id\right]
\left(
\textcolor{red}{\RN{1}_{6,5} \RN{1}_{6,4}} \RN{1}_{6,3} \textcolor{red}{\RN{1}_{6,2} \RN{1}_{6,1}} \right)\,.
\non 
\end{align}
This circuit can be used to prepare the 6-qubit Dicke state $|D^6(6-l,l) \rangle = U_6\, |\re(6-l,l)\rangle$ from the initial state $|\re(6-l,l)\rangle$ for any $l \in \{1, \dots, 5\}$. For the particular case $l=3$, the gates in \textcolor{red}{red} are redundant and
can therefore be removed, as explained below.

\subsection{Simplifying the circuit}\label{sec:simplify2}

For a Dicke state $|D^n(n-l,l) \rangle$ with a {\it given} (fixed) value of $l$, some of the gates in the above construction  \eqref{DickeOp}, \eqref{Uresult}, \eqref{W2explicit} are redundant, and can therefore be removed. We now prune away these redundant gates in order to obtain a simplified operator $\mathcal{U}_n(n-l,l)$ in terms of corresponding simplified operators $\mathcal{W}_m(n-l,l)$, such that
\begin{equation}
    \mathcal{U}_n(n-l,l)\, |\re(n-l,l)\rangle=|D^n(n-l,l)\rangle\,,
    \label{Dicke2simple}
\end{equation}
which are customized for a fixed value of $l$. 

We begin by considering how the right-most factor in \eqref{Uresult}, $W_n$, acts on $|\re(n-l,l)\rangle$ for a fixed $l$. The first property in \eqref{Iprops} implies
that $n-2$ factors in the product \eqref{W2explicit} can be removed, simplifying to $\mathcal{W}_n(n-l,l)=\RN{1}_{n,l}$. For example,
for $l=3$ in \eqref{U6example}, we can remove gates $\RN{1}_{6,1}, \RN{1}_{6,2}, 
\RN{1}_{6,4}, \RN{1}_{6,5}$ in $W_6$.

We next consider how $W_{n-1}\otimes\id$ acts on $\RN{1}_{n,l}\,|\re(n-l,l)\rangle$. Rewriting \eqref{W2explicit} as 
\begin{equation}
W_{n-1} = 
\left(\overset{\curvearrowleft}{\prod_{l'=l+1}^{n-2}}
\RN{1}_{n-1,l'} \right)
\RN{1}_{n-1,l}\, \RN{1}_{n-1,l-1}
\left(\overset{\curvearrowleft}{\prod_{l'=1}^{l-2}}
\RN{1}_{n-1,l'} \right)\,,
\end{equation}
we find that all the terms in the right product can be removed, as their controls are in qubit positions between and including $1$ and $l-1$, where the qubits take the value of $|1\rangle$. The terms in the left product can also be seen to leave the state invariant, and can therefore also be removed. 
Thus, the factor $W_{n-1}$ in \eqref{Uresult} can be simplified to
$\mathcal{W}_{n-1}(n-l,l)=\RN{1}_{n-1,l}\, \RN{1}_{n-1,l-1}$. 
For example,  for $l=3$ in \eqref{U6example}, the gates $\RN{1}_{5,1}$ and $\RN{1}_{5,4}$ in $W_5$ can be removed. 

Similar analysis can be done on the general $W_m$ factors in the product \eqref{Uresult}, leading to  
\begin{equation}
\mathcal{U}_n(n-l,l) =  \overset{\curvearrowright}{\prod_{m=2}^{n}} 
\left(\mathcal{W}_m(n-l,l) \otimes  \id^{\otimes(n-m)} \right)\,,
\label{Uresult2}
\end{equation}
where
\begin{equation}
\mathcal{W}_m(n-l,l)=
\overset{\curvearrowleft}{\prod\limits_{l'={\rm max}(l+m-n,1)}^{{\rm min}(l,m-1)}}
\,\RN{1}_{m,l'} \,.
\label{W2explicit2}
\end{equation}
The number of $\RN{1}$-gates in $\mathcal{U}_n(n-l,l)$ is given by 
\begin{equation}
    N^{\RN{1}}_n(l) = \sum_{m=2}^n \left[1 +
    {\rm min}(l,m-1) - {\rm max}(l+m-n,1) \right] \,,
\end{equation}
which satisfies $N^{\RN{1}}_n(l)=N^{\RN{1}}_n(n-l)$, and $N^{\RN{1}}_n(l) \sim l n$ for $l \ll n$. Hence, the circuit $\mathcal{U}_n(n-l,l)$ has size ${\cal O}({\rm min}(l,n-l) \cdot n)$, as in \cite{Bartschi2019}.

Cirq code that implements the qubit Dicke state constructions given by \eqref{DickeOp}, \eqref{Uresult}, \eqref{W2explicit}
as well as by \eqref{Dicke2simple}-\eqref{W2explicit2} is included in the Supplementary Material.

\section{The case $d=3$}\label{sec:d3} 

We now consider the case $d=3$ (qutrits). The defining relation for the $W$ operator \eqref{Wop} with $n=m$ now reduces to 
\begin{align}
W_m\, &|0\rangle^{\otimes k_0}|1\rangle^{\otimes k_{1}}
|2\rangle^{\otimes k_{2}} = \sqrt{\frac{k_{0}}{m}}\, |0\rangle^{\otimes 
(k_{0}-1)}|1\rangle^{\otimes k_{1}}
|2\rangle^{\otimes k_{2}} |0\rangle  \label{Wopd3} \\
&+ \sqrt{\frac{k_{1}}{m}}\, |0\rangle^{\otimes k_{0}}
|1\rangle^{\otimes (k_{1}-1)}
|2\rangle^{\otimes k_{2}} |1\rangle
+\sqrt{\frac{k_{2}}{m}}\, |0\rangle^{\otimes k_{0}}
|1\rangle^{\otimes k_{1}}
|2\rangle^{\otimes k_{2}}  \,, \qquad 
k_{0} + k_{1} + k_{2} = m \,. \non
\end{align}

\subsection{Elementary qutrit gates}\label{sec:gates}

We shall see that the $W$ operators can be decomposed entirely in 
terms of certain NOT gates, $R^{y}$ rotation gates, and controlled versions thereof.
Following \cite{Di:2011}\footnote{See also \cite{Wang:2020} and 
references therein.}, 
we denote by $X^{(ij)}$ the (1-qutrit) NOT 
gate that performs the interchange $|i\rangle \leftrightarrow 
|j\rangle$ and leaves unchanged the remaining basis vector, where $i, 
j \in \{0, 1, 2\}$ and $i<j$; that is,
\begin{align}
X^{(01)} |0\rangle &=  |1\rangle\,, \qquad X^{(01)} |1\rangle = 
|0\rangle \,, \qquad X^{(01)} |2\rangle = 
|2\rangle \,, \non \\
X^{(02)} |0\rangle &=  |2\rangle\,, \qquad X^{(02)} |2\rangle = 
|0\rangle \,, \qquad X^{(02)} |1\rangle = 
|1\rangle \,, \non \\
X^{(12)} |1\rangle &=  |2\rangle\,, \qquad X^{(12)} |2\rangle = 
|1\rangle \,, \qquad X^{(12)} |0\rangle = 
|0\rangle \,.
\label{NOTgates}
\end{align}
We similarly denote by $R^{(ij)}(\theta)$ the (1-qutrit) gate that 
performs an $R^{y}(\theta)$ rotation in the subspace spanned by $|i\rangle$ 
and $|j\rangle$; hence,
\begin{align}
R^{(ij)}(\theta) |i \rangle &= \cos(\theta/2) |i \rangle + 
\sin(\theta/2) |j \rangle \,, \non \\
R^{(ij)}(\theta) |j \rangle &= -\sin(\theta/2) |i \rangle + 
\cos(\theta/2) |j \rangle \,,
\label{Rgates}
\end{align}
with $(i,j) \in \{(0,1), (0,2), (1,2) \}$.

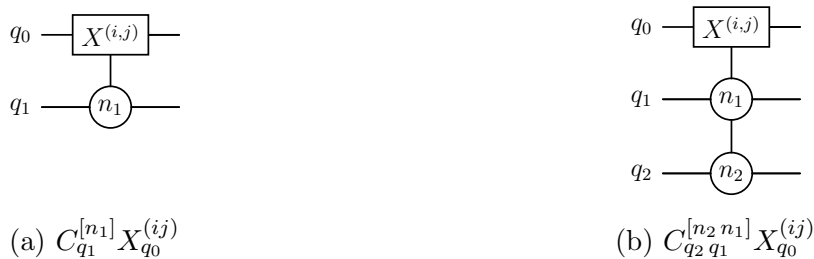
\begin{figure}[htb]
	\centering
	\begin{subfigure}{0.5\textwidth}
      \centering
\begin{adjustbox}{width=0.3\textwidth, raise=4em}
\begin{quantikz}
       \lstick{$q_{0}$} & \gate{X^{(i,j)}} \vqw{1} & \qw \\
       \lstick{$q_{1}$} & \gateO{n_{1}} & \qw 
       \end{quantikz}
\end{adjustbox}
	  \caption{$C^{[n_{1}]}_{q_{1}}X^{(ij)}_{q_{0}}$}
	  \label{fig:CX}
    \end{subfigure}%
    \begin{subfigure}{0.5\textwidth}
        \centering
\begin{adjustbox}{width=0.3\textwidth}
\begin{quantikz}
        \lstick{$q_{0}$} & \gate{X^{(i,j)}} \vqw{1} & \qw \\
        \lstick{$q_{1}$} & \gateO{n_{1}} \vqw{1} & \qw \\
        \lstick{$q_{2}$} & \gateO{n_{2}} & \qw 
        \end{quantikz}
\end{adjustbox}
	    \caption{$C^{[n_{2}\, n_{1}]}_{q_{2}\, q_{1}}X^{(ij)}_{q_{0}}$}
        \label{fig:CCX}
	 \end{subfigure}
\caption{Circuit diagrams for qutrit controlled-$X^{(ij)}$ gates}
\end{figure}

We denote by $C^{[n_{1}]}_{q_{1}}X^{(ij)}_{q_{0}}$ the (2-qutrit) controlled-$X^{(ij)}$ 
gate, which acts as $X^{(ij)}$ on the ``target'' qutrit in vector 
space $q_{0}$ if the 
``control'' qutrit in vector space $q_{1}$ is in the state $|n_{1}\rangle$, and otherwise 
acts as the identity operator. That is,
\begin{equation}
C^{[n_{1}]}_{q_{1}}X^{(ij)}_{q_{0}}\, |x_{1}\rangle_{q_{1}} 
|x_{0}\rangle_{q_{0}} = 
\begin{cases}
	|x_{1}\rangle_{q_{1}}  X^{(ij)} |x_{0}\rangle_{q_{0}} & \text{if} 
	\quad x_{1} = n_{1} \\
	|x_{1}\rangle_{q_{1}}  |x_{0}\rangle_{q_{0}} & \text{if} \quad 
	x_{1} \ne n_{1} 	
\end{cases} \,,
\label{CX}
\end{equation}
where $x_{0}, x_{1}, n_{1} \in \{0, 1, 2\}$, and $q_{0}, q_{1} \in 
\{0, 1, \ldots, n-1\}$. The corresponding circuit 
diagram is shown in Fig. \ref{fig:CX}.

\noindent
Similarly, $C^{[n_{2}\, n_{1}]}_{q_{2}\, q_{1}}X^{(ij)}_{q_{0}}$ denotes the (3-qutrit) 
double-controlled-$X^{(ij)}$ gate, with control qutrits in vector 
spaces $q_{1}$ and $q_{2}$, which must be in the states $|n_{1}\rangle$ and 
$|n_{2}\rangle$, respectively, in 
order for the gate to act nontrivially on the target qutrit in 
vector space $q_{0}$, see Fig. \ref{fig:CCX}; and similarly for 
higher multiple-controlled-$X^{(ij)}$ gates. Controlled $R^{y}$ rotation gates are defined in a similar 
way, with $X^{(ij)}$ replaced by $R^{(ij)}(\theta)$. 

Matrix representations of these gates and further notational details are presented in Appendix 
\ref{sec:matrices}. 

We now proceed in Sections \ref{sec:special} and \ref{sec:generic}
to use these gates to explicitly construct a circuit that implements the $W$ operator \eqref{Wopd3}.

\subsection{Special case}\label{sec:special}

Let us begin with the simpler special case that exactly one of the $k_j$'s is zero, i.e. either
\begin{equation}
k_{0}=0\,, \quad\text{ or  }\quad k_{1}=0\,, \quad\text{ or  }\quad k_{2}=0\,,
\label{specialcase}
\end{equation}
in which case \eqref{Wopd3} takes the form
\begin{equation}
W_m\, |i_0\rangle^{(m-l)}|i_1\rangle^{\otimes l}
= \sqrt{\frac{m-l}{m}}\, |i_0\rangle^{\otimes 
(m-l-1)}|i_1\rangle^{\otimes l}|i_0\rangle
+ \sqrt{\frac{l}{m}}\, |i_0\rangle^{\otimes 
(m-l)}|i_1\rangle^{\otimes l}  \,, 
\label{Wopd3special}
\end{equation}
where $i_0<i_1$; there are 3 such possibilities, namely 
$(i_0,i_1) \in \{ (0,1), (0,2), (1,2) \}$. Let us denote by $\widetilde{W}_m$
the restriction of $W_m$ to this special case \eqref{specialcase}, \eqref{Wopd3special}. 

We observe that $\widetilde{W}_m$ acts formally similarly to the 
qubit operator \eqref{Wop2}, except the latter involves only the single possibility $(i_0,i_1)=(0,1)$. We therefore introduce a qutrit operator $V_{m,l}^{(i_0,i_1)}$, similar to the qubit operator $I_{m,l}$ \eqref{Iaction}, that performs the mapping 
\begin{equation}
V_{m,l}^{(i_0,i_1)}:\quad |i_0\rangle_l\, |i_1\rangle_{l-1}\, 
|i_1\rangle_0 \mapsto
\sqrt{\frac{m-l}{m}}\, 
|i_1\rangle_l\, |i_1\rangle_{l-1}\, 
|i_0\rangle_0 +
\sqrt{\frac{l}{m}}\, 
|i_0\rangle_l\, 
|i_1\rangle_{l-1}\, 
|i_1\rangle_0\,,
\label{Vaction}
\end{equation}
and otherwise acts as identity (as long as the $0$th qutrit is in the state $|i_1\rangle$). For $l=1$, the middle qutrits in \eqref{Vaction} are omitted.
The operator $\widetilde{W}_m$ is then given, similarly to \eqref{W2explicit}, by
\begin{equation}
    \widetilde{W}_m = \overset{\curvearrowleft}{\prod_{l=1}^{m-1}}
    \RN{1}_{m,l} \,, \qquad
\RN{1}_{m,l} = V_{m,l}^{(1,2)}\, V_{m,l}^{(0,2)}\, V_{m,l}^{(0,1)}\,,
\label{W3special}
\end{equation}
where the order of the $V_{m,l}$'s in $\RN{1}_{m,l}$ is arbitrary. The circuit diagram for $V_{m,l}^{(i_0,i_1)}$ with $2 \le l \le 
m-2$ is given by Fig. \ref{fig:V3gen}, cf. Fig. \ref{fig:Iops}. For the edge cases $l=1$ and $l=m-1$, the corresponding circuit diagrams can be obtained from limits of Fig. \ref{fig:V3gen}, and
are given by Figs. \ref{fig:V3a} and \ref{fig:V3c}, respectively.
The control with $i_0>0$ is defined as an $i_0$ control that is present only if $i_0>0$; its role is to ensure for the case $(i_0,i_1)=(1,2)$ that the input state indeed consists only of $|1\rangle$'s and $|2\rangle$'s. Hence, 
$\RN{1}_{m,l}$ leaves invariant any generic (i.e., {\it not} special) initial state,
\begin{equation}
    \RN{1}_{m,l} \left( |0\rangle^{\otimes k_0} |1\rangle^{\otimes k_1} |2\rangle^{\otimes k_2}\right) = \left( |0\rangle^{\otimes k_0} |1\rangle^{\otimes k_1} |2\rangle^{\otimes k_2}\right) \qquad k_0, k_1, k_2 \ne 0 \,.
\end{equation}

\begin{figure}[htb]
	\centering
\begin{adjustbox}{width=0.4\textwidth}
\begin{quantikz}
\lstick{$0$} & \gateO{i_1} \vqw{1} & \gate{R^{(i_0,i_1)}(\theta)} 
\vqw{1}   & \gateO{i_1} \vqw{1} & \qw \\
\vdots &\vqw{1}&\vqw{1}&\vqw{1}& \\
\lstick{$l-1$} &\qw \vqw{1} & \gateO{i_1} \vqw{1} &\qw \vqw{1}  & \qw \\
\lstick{$l$} & \gate{X^{(i_0,i_1)}} \vqw{-1} & \gateO{i_1} \vqw{1} & 
\gate{X^{(i_0,i_1)}} \vqw{-1}    & \qw \\
\vdots &&&&\\
\lstick{$m-1$}&\qw &  \gateO{i_0>0}\vqw{-1}  &\qw &\qw   
\end{quantikz}
\end{adjustbox}
\caption{Circuit diagram for $V_{m,l}^{(i_0,i_1)}$ with $2 \le l \le 
m-2$, $m\ge4$, and $\theta$ in \eqref{ugate}}
\label{fig:V3gen}
\end{figure}	

\begin{figure}[htb]
	\centering
	\begin{subfigure}{0.5\textwidth}
      \centering
\begin{adjustbox}{width=0.8\textwidth, raise=7em}
\begin{quantikz}
\lstick{$0$} & \gateO{i_1} \vqw{1} & \gate{R^{(i_0,i_1)}(\theta)} 
\vqw{1}   & \gateO{i_1} \vqw{1} & \qw \\
\lstick{$l=1$} & \gate{X^{(i_0,i_1)}} \vqw{-1} & \gateO{i_1}  & 
\gate{X^{(i_0,i_1)}} \vqw{-1}   & \qw \\
\end{quantikz}
\end{adjustbox}
\caption{$V_{m,l}^{(i_0,i_1)}$ with $l=1$ and $m=2$}
\label{fig:I1} 
    \end{subfigure}%
    \begin{subfigure}{0.5\textwidth}
        \centering
\begin{adjustbox}{width=0.9\textwidth}
\begin{quantikz}
\lstick{$0$} & \gateO{i_1} \vqw{1} & \gate{R^{(i_0,i_1)}(\theta)} 
\vqw{1}   & \gateO{i_1} \vqw{1} & \qw \\
\lstick{$l=1$} & \gate{X^{(i_0,i_1)}} \vqw{-1} & \gateO{i_1} \vqw{1} & 
\gate{X^{(i_0,i_1)}} \vqw{-1}    & \qw \\
\vdots &&&&\\
\lstick{$m-1$}&\qw &  \gateO{i_0>0}\vqw{-1}  &\qw &\qw   
\end{quantikz}
\end{adjustbox}
\caption{$V_{m,l}^{(i_0,i_1)}$ with $l=1$ and $m>2$}
\label{fig:Il}
	 \end{subfigure}	
\caption{Circuit diagrams for $V_{m,l}^{(i_0,i_1)}$ with $l=1$ and $\theta$ in \eqref{ugate}}
\label{fig:V3a}
\end{figure}
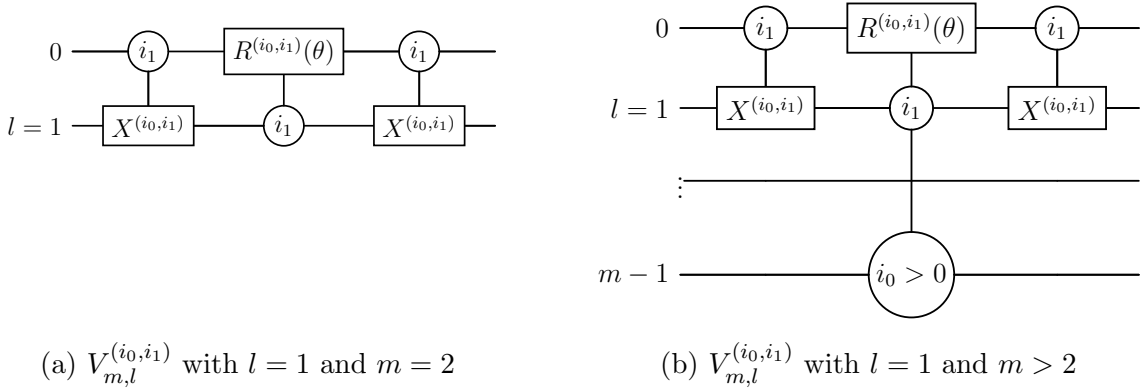

\begin{figure}[htb]
	\centering
\begin{adjustbox}{width=0.4\textwidth}
\begin{quantikz}
\lstick{$0$} & \gateO{i_1} \vqw{1} & \gate{R^{(i_0,i_1)}(\theta)} 
\vqw{1}   & \gateO{i_1} \vqw{1} & \qw \\
\vdots &\vqw{1}&\vqw{1}&\vqw{1}& \\
\lstick{$m-2$} &\qw \vqw{1} & \gateO{i_1} \vqw{1} &\qw \vqw{1}  & \qw \\
\lstick{$l=m-1$} & \gate{X^{(i_0,i_1)}} \vqw{-1} & \gateO{i_1}  & 
\gate{X^{(i_0,i_1)}} \vqw{-1}    & \qw \\  
\end{quantikz}
\end{adjustbox}
\caption{Circuit diagram for $V_{m,l}^{(i_0,i_1)}$ with $l=
m-1$, $m\ge3$, and $\theta$ in \eqref{ugate}}
\label{fig:V3c}
\end{figure}

\subsection{Generic case}\label{sec:generic}

In Sec. \ref{sec:special} we focused on the special case that exactly one of the $k_j$'s is zero \eqref{specialcase}, for which case $W_m$ 
generates only 2 terms \eqref{Wopd3special}, and therefore only one 
rotation angle $(\theta)$ is necessary.
Let us now consider the generic case that {\em all of the $k_j$'s are 
nonzero}, for which case $W_m$ generates 3 terms \eqref{Wopd3},
and therefore two rotation angles $(\theta_{1}, \theta_{2})$ are 
necessary.
Let us denote by $\widetilde{\widetilde{W}}_m$
the restriction of $W_m$ to this generic case. 
We will see that $\widetilde{\widetilde{W}}_m$ is given by a 
product of operators $\RN{2}$ depending on wire labels $l_{2} \in \{1, 2, \ldots\}$ and $l_{1} \in 
\{l_{2}+1\,, l_{2}+2\,, \ldots \}$, where  
\begin{equation}
l_{2}=k_{2}\,, \qquad l_{1}= k_{1} + k_{2} \,.
\label{l1l2ks}	
\end{equation}
Thus, $l_{2}-1$ is the wire in the 
initial state $|\re(\vec k)\rangle$ with the ``last'' $|2\rangle$, 
and $l_{1}-1$ is the wire with the ``last'' $|1\rangle$, going from right to 
left:
\begin{equation}
|0\rangle_{m-1} 
\cdots 
|0\rangle_{l_{1}}
|1\rangle_{l_{1}-1}
\cdots 
|1\rangle_{l_{2}}
|2\rangle_{l_{2}-1} 
\cdots 
|2\rangle_{0} \,.
\label{l1l2}
\end{equation}

We introduce the operator $\RN{2}_{m,l_1,l_2}$ acting on the $l_1$th, $(l_1-1)$th, $l_2$th, $(l_2-1)$th, and $0$th qutrit, which performs the transformation
\begin{align}
\RN{2}_{m,l_1,l_2}:\quad 
|0\rangle_{l_{1}} |1\rangle_{l_{1}-1}
|1\rangle_{l_{2}} |2\rangle_{l_{2}-1}
|2\rangle_{0} & \mapsto 
\cos(\theta_{1}/2)
|0\rangle_{l_{1}} |1\rangle_{l_{1}-1}
|1\rangle_{l_{2}} |2\rangle_{l_{2}-1}
|2\rangle_{0} \non \\
&- \sin(\theta_{1}/2)\cos(\theta_{2}/2)
|0\rangle_{l_{1}} |1\rangle_{l_{1}-1} 
|2\rangle_{l_{2}} |2\rangle_{l_{2}-1}
|1\rangle_{0} \non\\
&+ \sin(\theta_{1}/2)\sin(\theta_{2}/2)
|1\rangle_{l_{1}} |1\rangle_{l_{1}-1} 
|2\rangle_{l_{2}} |2\rangle_{l_{2}-1}
|0\rangle_{0} \,,
\label{stateIId}
\end{align}
and otherwise acts as identity (as long as the $0$th qutrit is in the state $|2\rangle$, which is always the case for generic input states in \eqref{Wopd3}). 
For $l_2=1$, the next-to-rightmost qutrits in \eqref{stateIId} are omitted; and for 
$l_1=l_2+1$, the next-to-leftmost qutrits in \eqref{stateIId} are omitted. We demand
\begin{align}
\cos(\theta_{1}/2) &= \sqrt{\frac{l_{2}}{m}}\,, \qquad
\sin(\theta_{1}/2)\cos(\theta_{2}/2) = - \sqrt{\frac{l_{1}-l_{2}}{m}} 
\,, \non \\
& \sin(\theta_{1}/2)\sin(\theta_{2}/2) = \sqrt{\frac{m-l_{1}}{m}} \,,
\label{demand}
\end{align}
in order to match with \eqref{Wopd3} and  \eqref{l1l2ks}. We therefore 
assign to the rotation angles the values
\begin{equation}
\theta_{1} = -2 \arccos\left(\sqrt{\frac{l_{2}}{m}}\right)\,, \qquad 
\theta_{2} = -2 \arccos\left(\sqrt{\frac{l_{1}-l_{2}}{m-l_{2}}}\right) \,.
\label{angles}
\end{equation}

The operator $\RN{2}_{m,l_1,l_2}$ in \eqref{stateIId}
(with $l_2 + 1 < l_1 \le m-1$, $l_{2} > 1$, $m>4$) 
can be implemented by the circuit in Fig. \ref{fig:IIgen}. We note that $\RN{2}_{m,l_1,l_2}$ does not require the control $i_0>0$ on the $(m-1)$th wire (as is required for $V_{m,l}^{(i_0,i_1)}$) since $\RN{2}_{m,l_1,l_2}$ becomes activated only when the $(m-1)$th wire is in the state $|0\rangle$.

\begin{figure}[htb]
	\centering
\begin{adjustbox}{width=0.7\textwidth}
\begin{quantikz}
\lstick{$0$} & \gateO{2} \vqw{1} &  \gate{R^{(1,2)}(\theta_{1})} \vqw{1} & 
\gateO{2} \vqw{1} & \gateO{1} \vqw{1} &  \gate{R^{(0,1)}(\theta_{2})} \vqw{1} & \gateO{1} \vqw{1} & \qw \\
\vdots
&\vqw{1}&\vqw{1}&\vqw{1}&\vqw{1}&\vqw{1}&\vqw{1} & \\
\lstick{$l_{2}-1$} &\qw \vqw{1}  &\gateO{2}\vqw{1} &\qw\vqw{1} &\qw\vqw{1} 
&\gateO{2}\vqw{1} &\qw\vqw{1} &\qw \\
\lstick{$l_{2}$} & \gate{X^{(1,2)}} & \gateO{2} \vqw{1} & \gate{X^{(1,2)}}  &\qw \vqw{1}
 & \gateO{2} \vqw{1} & \qw \vqw{1} & \qw \\ 
\vdots
& &\vqw{1}& &\vqw{1}&\vqw{1}&\vqw{1} \\
\lstick{$l_{1}-1$} &\qw & \gateO{1} \vqw{1} & \qw & \qw\vqw{1} & 
\gateO{1} \vqw{1} & \qw \vqw{1} & \qw \\ 
\lstick{$l_{1}$} &\qw & \gateO{0} & \qw & \gate{X^{(0,1)}} & 
\gateO{1}  & \gate{X^{(0,1)}} & \qw \\ 
\vdots\\
\lstick{$m-1$} &\qw&\qw&\qw&\qw&\qw&\qw&\qw\\
\end{quantikz}
\end{adjustbox}
\caption{Circuit diagram for $\RN{2}_{m,l_{1}, l_{2}}$ with 
$l_2 + 1 < l_1 \le m-1$, $l_{2} > 1$, $m>4$, and 
$\theta_{1}\,, \theta_{2}$ in \eqref{angles}}
\label{fig:IIgen}
\end{figure}	

For the three types of edge cases: 
\begin{enumerate}[label=(\roman*)]
    \item $l_2=1$, $l_1=2$, $m>2$
    \item $l_2=1$, $2 < l_1 \le m-1$, $m>3$
    \item $l_2>1$, $l_1 = l_2+1 \le m-1$, $m>3$
\end{enumerate}
the corresponding circuit diagrams for $\RN{2}_{m,l_1,l_2}$ 
can be obtained from limits of Fig. \ref{fig:IIgen},
see Figs. \ref{fig:IIa}, \ref{fig:IIb}, \ref{fig:IIc}, respectively.

\begin{figure}[htb]
	\centering
\begin{adjustbox}{width=0.7\textwidth}
\begin{quantikz}
\lstick{$0$} & \gateO{2} \vqw{1} &  \gate{R^{(1,2)}(\theta_{1})} \vqw{1} & 
\gateO{2} \vqw{1} & \gateO{1} \vqw{1} &  \gate{R^{(0,1)}(\theta_{2})} \vqw{1} & \gateO{1} \vqw{1} & \qw \\
\lstick{$l_{2}=1$} & \gate{X^{(1,2)}} & \gateO{2} \vqw{1} & \gate{X^{(1,2)}}  &\qw \vqw{1}
 & \gateO{2} \vqw{1} & \qw \vqw{1} & \qw \\ 
\lstick{$l_{1}=2$} &\qw & \gateO{0} & \qw & \gate{X^{(0,1)}} & 
\gateO{1}  & \gate{X^{(0,1)}} & \qw \\ 
\vdots\\
\lstick{$m-1$} &\qw&\qw&\qw&\qw&\qw&\qw&\qw\\
\end{quantikz}
\end{adjustbox}
\caption{Circuit diagram for $\RN{2}_{m,l_{1}, l_{2}}$ with 
$l_2=1$, $l_1=2$, $m>2$, and 
$\theta_{1}\,, \theta_{2}$ in \eqref{angles}}
\label{fig:IIa}
\end{figure}	

\begin{figure}[htb]
	\centering
\begin{adjustbox}{width=0.7\textwidth}
\begin{quantikz}
\lstick{$0$} & \gateO{2} \vqw{1} &  \gate{R^{(1,2)}(\theta_{1})} \vqw{1} & 
\gateO{2} \vqw{1} & \gateO{1} \vqw{1} &  \gate{R^{(0,1)}(\theta_{2})} \vqw{1} & \gateO{1} \vqw{1} & \qw \\
\lstick{$l_{2}=1$} & \gate{X^{(1,2)}} & \gateO{2} \vqw{1} & \gate{X^{(1,2)}}  &\qw \vqw{1}
 & \gateO{2} \vqw{1} & \qw \vqw{1} & \qw \\ 
\vdots
& &\vqw{1}& &\vqw{1}&\vqw{1}&\vqw{1} \\
\lstick{$l_{1}-1$} &\qw & \gateO{1} \vqw{1} & \qw & \qw\vqw{1} & 
\gateO{1} \vqw{1} & \qw \vqw{1} & \qw \\ 
\lstick{$l_{1}$} &\qw & \gateO{0} & \qw & \gate{X^{(0,1)}} & 
\gateO{1}  & \gate{X^{(0,1)}} & \qw \\ 
\vdots\\
\lstick{$m-1$} &\qw&\qw&\qw&\qw&\qw&\qw&\qw\\
\end{quantikz}
\end{adjustbox}
\caption{Circuit diagram for $\RN{2}_{m,l_{1}, l_{2}}$ with 
$l_2=1$, $2 < l_1 \le m-1$, $m>3$, and 
$\theta_{1}\,, \theta_{2}$ in \eqref{angles}}
\label{fig:IIb}
\end{figure}	

\begin{figure}[htb]
	\centering
\begin{adjustbox}{width=0.7\textwidth}
\begin{quantikz}
\lstick{$0$} & \gateO{2} \vqw{1} &  \gate{R^{(1,2)}(\theta_{1})} \vqw{1} & 
\gateO{2} \vqw{1} & \gateO{1} \vqw{1} &  \gate{R^{(0,1)}(\theta_{2})} \vqw{1} & \gateO{1} \vqw{1} & \qw \\
\vdots
&\vqw{1}&\vqw{1}&\vqw{1}&\vqw{1}&\vqw{1}&\vqw{1} & \\
\lstick{$l_{2}-1$} &\qw \vqw{1}  &\gateO{2}\vqw{1} &\qw\vqw{1} &\qw\vqw{1} 
&\gateO{2}\vqw{1} &\qw\vqw{1} &\qw \\
\lstick{$l_{2}$} & \gate{X^{(1,2)}} & \gateO{2} \vqw{1} & \gate{X^{(1,2)}}  &\qw \vqw{1}
 & \gateO{2} \vqw{1} & \qw \vqw{1} & \qw \\ 
\lstick{$l_{1}=l_2+1$} &\qw & \gateO{0} & \qw & \gate{X^{(0,1)}} & 
\gateO{1}  & \gate{X^{(0,1)}} & \qw \\ 
\vdots\\
\lstick{$m-1$} &\qw&\qw&\qw&\qw&\qw&\qw&\qw\\
\end{quantikz}
\end{adjustbox}
\caption{Circuit diagram for $\RN{2}_{m,l_{1}, l_{2}}$ with 
$l_2>1$, $l_1 = l_2+1 \le m-1$, $m>3$, and 
$\theta_{1}\,, \theta_{2}$ in \eqref{angles}}
\label{fig:IIc}
\end{figure}	

We note that these operators satisfy the following properties
\begin{align}
    \RN{2}_{m,l_1',l'_2}\,|\re(m-l_1, l_1-l_2,l_2)\rangle &= |\re(m-l_1, l_1-l_2,l_2)\rangle  \non\\
    &\qquad  \text{ if } \quad l_1' \ne l_1  \quad \text{ or } \quad 
   l_2' \ne l_2   \,, \nonumber\\
   \RN{2}_{m,l_1',l'_2}\,
   \left[  \RN{2}_{m,l_1,l_2}\,|\re(m-l_1, l_1-l_2,l_2)\rangle\right]
     &= \RN{2}_{m,l_1,l_2}\,|\re(m-l_1, l_1-l_2,l_2)\rangle \non\\
     &\qquad \text{ if } \quad l_1' > l_1 \quad \text{ or } \quad 
   l_2' > l_2   \,,  \nonumber\\
   \RN{2}_{m,l_1,l_2}\,
   \left[  \RN{1}_{m,l} \left( |i\rangle^{\otimes(m-l)} |j\rangle^{\otimes l} \right)  \right]
     &= \RN{1}_{m,l} \left( |i\rangle^{\otimes(m-l)} |j\rangle^{\otimes l} \right) \non\\
     &\qquad \text{ if } \quad  l_1 > l_2 \quad \text{ and } \quad  i<j   \,.
   \label{IIprops}
\end{align}
The desired $\widetilde{\widetilde{W}}_m$ operator is therefore given by an ordered product
of all possible $\RN{2}$-operators
\begin{equation}
\widetilde{\widetilde{W}}_m = \overset{\curvearrowleft}
{\prod_{l_2=1}^{m-2}\prod_{l_1=l_2+1}^{m-1}}
\RN{2}_{m,l_1, l_2} \,,
\label{W3generic}
\end{equation}
where either $l_2$ or $l_1$ increases from right to left. For example,
\begin{equation}
    \widetilde{\widetilde{W}}_4 = \RN{2}_{4,3,2}\, \RN{2}_{4,3,1}\, \RN{2}_{4,2,1} \,.
\end{equation}

\subsection{Summarizing}\label{sec:together}

For the general $d=3$ case (with no conditions on $\vec k$, apart from 
$k_{0} + k_{1} + k_{2} = m$), we obtain our result for
an operator $W_{m}$ (independent of $\vec k$) that satisfies \eqref{Wopd3}, namely
\begin{equation}
W_{m} = \widetilde{\widetilde{W}}_{m}\, \widetilde{W}_{m} \,,
\label{W3final}
\end{equation}
where $\widetilde{W}_{m}$ is given by \eqref{W3special}, and 
$\widetilde{\widetilde{W}}_{m}$ is given by \eqref{W3generic}.

The size and depth of the qutrit circuit $U_n$ is ${\cal O}(n^3)$, see \eqref{size} below. We note that
the multi-controlled qutrit $R^y$ gates in this circuit can be decomposed into elementary 1-qutrit and 2-qutrit gates in the same way as for corresponding multi-controlled {\it qubit} $R^y$ gates, since we use the naive embeddings $SU(2) \subset SU(3)$ \eqref{Rij}. The number of 2-qubit gates in the decomposition of multi-controlled qubit $R^y$ gates, as provided by cirq, is displayed in Table \ref{table:decomp}.

\begin{table}[htb]
\centering
\begin{tabular}{c|c}
number of controls & number of 2-qubit gates \\
\hline
1 & 2 \\
2 & 8 \\  
3 & 22 \\
4 & 50 \\
\end{tabular}
\caption{The number of 2-qubit gates in the decomposition of multi-controlled qubit $R^y$ gates}\label{table:decomp}
\end{table}

The initial state $|\re(\vec k) \rangle$ (recall Eqs. \eqref{initialstate} and 
\eqref{DickeOp}) with $d=3$ is readily 
constructed by applying $X^{(01)}$ and $X^{(02)}$ gates to the 
all-$|0\rangle$ $n$-qutrit state 
\begin{equation}
|\re(\vec k) \rangle	=  |0\rangle^{\otimes k_{0}}|1\rangle^{\otimes 
k_{1}} |2\rangle^{\otimes k_{2}} =  \id^{\otimes k_{0}} \otimes
(X^{(01)})^{\otimes k_{1}} \otimes 
(X^{(02)})^{\otimes k_{2}} |0\rangle^{\otimes n} \,.
\label{qutritinitstate}
\end{equation}

As a simple example, the 3-qutrit Dicke state $|D^{3}(1,1,1)\rangle$ is obtained by 
\begin{equation}
|D^{3}(1,1,1)\rangle = 	U_{3}\, |0\rangle|1\rangle |2\rangle\,, 
\label{D111}
\end{equation}
where
\begin{equation}
U_{3} =(W_{2} \otimes \id)\, W_{3} 
=(\RN{1}_{2,1} \otimes \id)\, 
\left( \RN{2}_{3,2,1}\, \textcolor{red}{\RN{1}_{3,2}}\, \RN{1}_{3,1} \right)\,,
\label{U3ex}
\end{equation}
see Eqs. \eqref{DickeOp}, \eqref{Uresult}, \eqref{W3special}, \eqref{W3generic}, \eqref{W3final}.
The $\RN{1}_{m,l}$'s are given in terms of 
$V_{m,l}^{(i_0,i_1)}$'s, see Eq. \eqref{W3special}, where the latter are given by Figs. \ref{fig:V3gen} and \ref{fig:V3a}; and $\RN{2}_{3,2,1}$ is given by Fig. \ref{fig:IIa}.

Similarly, the 4-qutrit state $|D^{4}(2,1,1)\rangle$ \eqref{d3example} is obtained by
\begin{equation}
|D^{4}(2,1,1)\rangle = 	U_{4}\, |0\rangle^{\otimes 2}|1\rangle |2\rangle\,,
\label{D211}
\end{equation}
with
\begin{align}
U_{4} &= \left(W_{2} \otimes \id^{\otimes 2} \right) \left(W_{3} \otimes \id \right) W_{4}\non \\
&=  \left( \RN{1}_{2,1} \otimes \id^{\otimes 2}  \right)
\left(\RN{2}_{3,2,1}\, \textcolor{red}{\RN{1}_{3,2}}\, \RN{1}_{3,1} \otimes \id \right) 
\left( \textcolor{red}{\RN{2}_{4,3,2}}\, \textcolor{red}{\RN{2}_{4,3,1}}\, \RN{2}_{4,2,1}\, 
\textcolor{red}{\RN{1}_{4,3}}\, \textcolor{red}{\RN{1}_{4,2}}\, \RN{1}_{4,1}\right) \,.
\label{U4ex}
\end{align}

The gates in \textcolor{red}{red} in Eqs. \eqref{U3ex} and \eqref{U4ex} are redundant (for generating the states $|D^{3}(1,1,1)\rangle$ and $|D^{4}(2,1,1)\rangle$, respectively)
and can therefore be removed, as explained below.

\subsection{Simplifying the circuit}\label{sec:simplify3}

The operator $U_n$ \eqref{Uresult}, with the $W$ operators given by \eqref{W3final},
generates the qutrit Dicke state $|D^n(\vec k)\rangle$ for any $\vec k$, see \eqref{DickeOp}. For a fixed $\vec k$, it is possible to prune away redundant gates, and therefore reduce the circuit size,
as we did in Sec. \ref{sec:simplify2} for $d=2$. We therefore now look for simplified operators $\mathcal{U}_n(n-l_1,l_1-l_2,l_2)$ and $\mathcal{W}_m(n-l_1,l_1-l_2,l_2)$, depending on given values $l_1$ and $l_2$ (which are related to $\vec k$ by \eqref{l1l2ks}), such that
\begin{align}
\mathcal{U}_n(n-l_1,l_1-l_2,l_2)\, |\re(n-l_1,l_1-l_2,l_2)\rangle
&=|D^n(n-l_1,l_1-l_2,l_2)\rangle\,, \non \\
\mathcal{U}_n(n-l_1,l_1-l_2,l_2) &=  \overset{\curvearrowright}{\prod_{m=2}^{n}} 
\left(\mathcal{W}_m(n-l_1,l_1-l_2,l_2) \otimes  \id^{\otimes(n-m)} \right)\,.
\label{Dicke3simple}
\end{align}
Setting as in \eqref{W3final}
\begin{equation}
\mathcal{W}_m(n-l_1,l_1-l_2,l_2) = 
\widetilde{\widetilde{\mathcal{W}}}_m(n-l_1,l_1-l_2,l_2)\,
\widetilde{\mathcal{W}}_m(n-l_1,l_1-l_2,l_2) \,,
\label{W3finalsimple}
\end{equation}
we conjecture that, similarly to the $d=2$ case \eqref{W2explicit2}, 
\begin{equation}
\widetilde{\widetilde{\mathcal{W}}}_m(n-l_1,l_1-l_2,l_2) 
= \overset{\curvearrowleft}
{\prod_{l'_2= {\rm max}(l_2+m-n,1)}^{{\rm min}(l_2,m-2)}
\prod_{l'_1= {\rm max}(l_1+m-n,l'_2+1)}^{{\rm min}(l_1,m-1)}}
\RN{2}_{m,l'_1, l'_2} \,,
\label{W3genericsimple}
\end{equation}
and
\begin{equation}
\widetilde{\mathcal{W}}_m(n-l_1,l_1-l_2,l_2) 
=
\overset{\curvearrowleft}{\prod_{l={\rm max}(\tilde{k}+m-n,1)}^{{\rm min}(\tilde{k},m-1)}}\RN{1}_{m, l} \,,
\label{W3specialsimple}
\end{equation}
where $\RN{1}_{m,l}$ is defined in \eqref{W3special}, and 
$\tilde{k}$ is defined (in terms of $k_0 = n-l_1$, $k_1=l_1-l_2$, and $k_2 = l_2$) by
\begin{align}
    \tilde{k} &=\begin{cases}
       k_2 & \text{if} \quad k_0 = 0 \\
    {\rm max}(k_1, k_2) & \text{if} \quad k_0 \ne 0  \\    
    \end{cases}  \,.
    \label{kkdefs}
\end{align}
We have not yet succeeded to prove the result \eqref{W3genericsimple}-\eqref{kkdefs}, which we found through experimentation.

Cirq code that implements the qutrit Dicke state constructions given by 
\eqref{DickeOp}, \eqref{Uresult}, \eqref{W3final}
as well as by \eqref{Dicke3simple}-\eqref{kkdefs} is included in the Supplementary Material.

The number of $\RN{1}$-gates and  $\RN{2}$-gates in $\mathcal{U}_n(n-l_1,l_1-l_2,l_2)$ is given by 
\begin{align}
    N^{\RN{1}}_n(l_1,l_2) &= \sum_{m=2}^n \left[1 +
    {\rm min}(\tilde{k},m-1) - {\rm max}(\tilde{k}+m-n,1) \right] \,, \non \\
    N^{\RN{2}}_n(l_1,l_2) &= \sum_{m=2}^n
    \sum_{l'_2= {\rm max}(l_2+m-n,1)}^{{\rm min}(l_2,m-2)}
    \left[1 +
    {\rm min}(l_1,m-1) - {\rm max}(l_1+m-n,l'_2+1) \right] \,,
\end{align}
respectively. For $l_1 \sim l_2 \equiv l \ll n$, we see that $N^{\RN{1}}_n(l_1,l_2) \sim l n$ and $N^{\RN{2}}_n(l_1,l_2) \sim l^2 n$. The size and depth of the simplified circuit $\mathcal{U}_n(n-l_1,l_1-l_2,l_2)$ is therefore ${\cal O}(l^2 n)$.

\section{General $d$}\label{sec:Wopgen} 

\begin{figure}[htb]
	\centering
\begin{adjustbox}{width=1.0\textwidth}
\begin{quantikz}
\lstick{$0$} & \gateO{i_{j-1}} \vqw{1} & \gate{R^{(i_{j-2},i_{j-1})}(\theta_{1})} \vqw{1}& 
\gateO{i_{j-1}} \vqw{1} & \gateO{i_{j-2}} \vqw{1} &  
\gate{R^{(i_{j-3},i_{j-2})}(\theta_{2})} \vqw{1} & \gateO{i_{j-2}} \vqw{1} & 
\qw &
\cdots\quad & \gateO{i_k} \vqw{1}&  
\gate{R^{(i_{k-1},i_k)}(\theta_{j-k})} \vqw{1}& 
\gateO{i_k} \vqw{1} & \qw  & \cdots\quad &  \gateO{i_1} \vqw{1} &  
\gate{R^{(i_0,i_1)}(\theta_{j-1})} \vqw{1} & \gateO{i_1} \vqw{1} & \qw \\
\vdots 
&\vqw{1}&\vqw{1}&\vqw{1}&\vqw{1}&\vqw{1}&\vqw{1}&&&\vqw{1}&\vqw{1}&\vqw{1}&&&\vqw{1}&\vqw{1}&\vqw{1}&\\
\lstick{$l_{j-1}-1$} &\qw\vqw{1} 
&\gateO{i_{j-1}}\vqw{1}&\qw\vqw{1}&\qw\vqw{1}&\gateO{i_{j-1}}\vqw{1}&\qw\vqw{1}&
\qw&\cdots\quad &\qw\vqw{1}&\gateO{i_{j-1}}\vqw{1}&\qw\vqw{1}&\qw&\cdots\quad&\qw\vqw{1}&\gateO{i_{j-1}}\vqw{1}&\qw\vqw{1}&\qw\\
\lstick{$l_{j-1}$} & \gate{X^{(i_{j-2},i_{j-1})}}
&\gateO{i_{j-1}}\vqw{1}&\gate{X^{(i_{j-2},i_{j-1})}}&\qw\vqw{1}&\gateO{i_{j-1}}\vqw{1}&\qw\vqw{1}&\qw&\cdots\quad 
&\qw\vqw{1}&\gateO{i_{j-1}}\vqw{1}&\qw\vqw{1}&\qw&\cdots\quad&\qw\vqw{1}&\gateO{i_{j-1}}\vqw{1}&\qw\vqw{1}&\qw\\
\vdots 
&&\vqw{1}&&\vqw{1}&\vqw{1}&\vqw{1}&&&\vqw{1}&\vqw{1}&\vqw{1}&&&\vqw{1}&\vqw{1}&\vqw{1}&\\
\lstick{$l_{j-2}-1$} &\qw 
&\gateO{i_{j-2}}\vqw{1}&\qw&\qw\vqw{1}&\gateO{i_{j-2}}\vqw{1}&\qw\vqw{1}&
\qw&\cdots\quad &\qw\vqw{1}&\gateO{i_{j-2}}\vqw{1}&\qw\vqw{1}&\qw&\cdots\quad&\qw\vqw{1}&\gateO{i_{j-2}}\vqw{1}&\qw\vqw{1}&\qw\\
\lstick{$l_{j-2}$} &\qw 
&\gateO{i_{j-3}}\vqw{1}&\qw&\gate{X^{(i_{j-3},i_{j-2})}}&\gateO{i_{j-2}}\vqw{1}&\gate{X^{(i_{j-3},i_{j-2})}}&
\qw&\cdots\quad 
&\qw\vqw{1}&\gateO{i_{j-2}}\vqw{1}&\qw\vqw{1}&\qw&\cdots\quad&\qw\vqw{1}&\gateO{i_{j-2}}\vqw{1}&\qw\vqw{1}&\qw\\
\vdots 
&&\vqw{1}&&&\vqw{1}&&&&\vqw{1}&\vqw{1}&\vqw{1}&&&\vqw{1}&\vqw{1}&\vqw{1}&\\
\lstick{$l_{k}-1$} &\qw 
&\gateO{i_k}\vqw{1}&\qw&\qw&\gateO{i_k}\vqw{1}&\qw&
\qw&\cdots\quad 
&\qw\vqw{1}&\gateO{i_k}\vqw{1}&\qw\vqw{1}&\qw&\cdots\quad&\qw\vqw{1}&\gateO{i_k}\vqw{1}&\qw\vqw{1}&\qw\\
\lstick{$l_{k}$} &\qw 
&\gateO{i_{k-1}}\vqw{1}&\qw&\qw&\gateO{i_{k-1}}\vqw{1}&\qw&
\qw&\cdots\quad 
&\gate{X^{(i_{k-1},i_k)}}&\gateO{i_k}\vqw{1}&\gate{X^{(i_{k-1},i_k)}}&\qw&\cdots\quad&\qw\vqw{1}&\gateO{i_k}\vqw{1}&\qw\vqw{1}&\qw\\
\vdots 
&&\vqw{1}&&&\vqw{1}&&&&&\vqw{1}&&&&\vqw{1}&\vqw{1}&\vqw{1}&\\
\lstick{$l_{1}-1$} &\qw 
&\gateO{i_1}\vqw{1}&\qw&\qw&\gateO{i_1}\vqw{1}&\qw&
\qw&\cdots\quad 
&\qw&\gateO{i_1}\vqw{1}&\qw&\qw&\cdots\quad&\qw\vqw{1}&\gateO{i_1}\vqw{1}&\qw\vqw{1}&\qw\\
\lstick{$l_{1}$} &\qw &\gateO{i_0}\vqw{2} &\qw&\qw&\gateO{i_0}\vqw{2} &\qw&
\qw&\cdots\quad 
&\qw&\gateO{i_0}\vqw{2} 
&\qw&\qw&\cdots\quad&\gate{X^{(i_0,i_1)}}&\gateO{i_1} \vqw{2}&\gate{X^{(i_0,i_1)}}&\qw\\
\vdots&&&&&&&&&&&&&&&&&\\
\lstick{$m-1$}&\qw&\gateO{i_0>0}\qw&\qw&\qw&\gateO{i_0>0}\qw&\qw&\qw&\cdots\quad&\qw&\gateO{i_0>0}\qw&\qw&\qw&\cdots\quad&\qw&\gateO{i_0>0}\qw&\qw&\qw\\
\end{quantikz}
\end{adjustbox}
\caption{Circuit diagram for 
$V^{(i_0, i_1, \dots, i_{j-1})}_{m, l_1, \ldots, l_{j-1}}$}
\label{fig:dgen}
\end{figure}
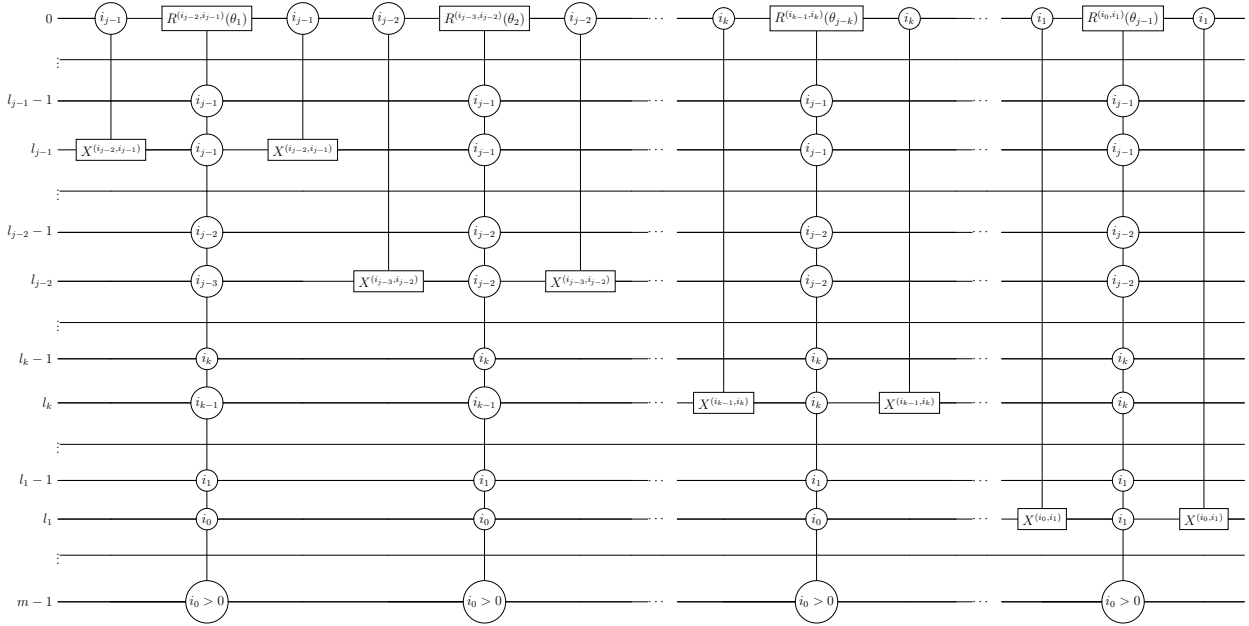

We now consider the decomposition of the $W$ operator \eqref{Wop} for general 
values of $d$ in terms of elementary qudit gates, which are defined 
similarly to the qutrit gates reviewed in Sec. \ref{sec:gates}. Exhibiting the $d$-dependence explicitly, Eqs. \eqref{DickeOp} and \eqref{Uresult} become
\begin{equation}
    U_n^{(d)}\, |\re(\vec{k})\rangle=|D^n(\vec k)\rangle\,,
\label{DickeOpd}
\end{equation}
and
\begin{equation}
U_n^{(d)} =  \overset{\curvearrowright}{\prod_{m=2}^{n}} 
\left(W_m^{(d)} \otimes  \id^{\otimes(n-m)} \right)\,.
\label{Uresultd}
\end{equation}
We find that $W_m^{(d)}$ is given by
\begin{equation}
W_m^{(d)} = \overset{\curvearrowleft}{\prod_{j=2}^{d}}
W_m^{(d,j)}	\,, 
\label{Wmd}
\end{equation}
where 
\begin{equation}
    W_m^{(d,j)} =  \overset{\curvearrowleft}
{\prod_{l_{j-1}=1}^{m-j+1} \prod_{l_{j-2}=l_{j-1}+1}^{m-j+2}\cdots 
\prod_{l_2=l_3+1}^{m-2} \prod_{l_1=l_2+1}^{m-1}}
\left[  \prod_{0 \le i_0 < i_1 < \dots < i_{j-1} \le d-1} V^{(i_0, i_1, \dots, i_{j-1})}_{m, l_1, \ldots, l_{j-1}}
\right] \,,
\label{Wmdj}
\end{equation}
and the circuit diagram for the operator $V^{(i_0, i_1, \dots, i_{j-1})}_{m, l_1, \ldots, l_{j-1}}$, with $0 \le i_0 < i_1 < \dots < i_{j-1} \le d-1$ and 
$1\le l_{j-1} < l_{j-2} < \dots < l_1 \le m-1$, is given by Fig. \ref{fig:dgen}. 
This operator has $j-1$ rotation angles $\theta_{1}, \ldots, 
\theta_{j-1}$, which can be determined in terms of the $l$'s and $m$ from the relations
\begin{align}
\cos(\theta_{1}/2)	&= \sqrt{\frac{l_{j-1}}{m}} \,, \non\\
\cos(\theta_{j-i}/2) \prod_{i'=1}^{j-i-1}(-\sin(\theta_{i'}/2)) &= 
\sqrt{\frac{l_{i}-l_{i+1}}{m}} \,, \qquad i = 1, \ldots, j-2  \,, \non\\
\prod_{i'=1}^{j-1}(-\sin(\theta_{i'}/2)) &= \sqrt{\frac{m-l_{1}}{m}} \,. 
\end{align}	
Circuit diagrams for edge cases can be obtained from suitable limits of Fig. \ref{fig:dgen}, as for $d=3$.
For a given value of $j$, the operators $V^{(i_0, i_1, \dots, i_{j-1})}_{m, l_1, \ldots, l_{j-1}}$ act nontrivially on states $|\re(\vec k) \rangle$ for which the number of nonzero $k_i$'s is $j$ (i.e., $j=d - \sum_{i=0}^{d-1} \delta_{k_i,0}$).

As a check on this result, let us count the number of $V$ operators in $W_m^{(d)}$ \eqref{Wmd}. The number of $V$ operators in the 
product within square brackets in \eqref{Wmdj} is given by ${d \choose j}$ (namely, the number of ways of choosing the $j$ integers $i_0, i_1, \ldots, i_{j-1}$ from the set of $d$ integers $\{0, 1, \ldots, d-1\}$.) Moreover, the number of possible values of $l_1, \ldots, l_{j-1}$ in \eqref{Wmdj} is given by ${m-1 \choose j-1}$. The number of $V$ operators in $W_m^{(d,j)}$ is therefore ${d \choose j}{m-1 \choose j-1}$. We conclude that the number of $V$ operators in $W_m^{(d)}$  is given by\footnote{For $j=1$, we have $W_m^{(d,1)} = \id$, the identity operator.}
\begin{equation}
    \sum_{j=1}^d {d \choose j}{m-1 \choose j-1} = {m+d-1 \choose d-1} \,.
\label{weakcomps}
\end{equation}
The result \eqref{weakcomps} is the number of weak $d$-compositions of $m$ \cite{Stanley2011}, 
i.e. the number of ways of writing $m$ as $\sum_{i=0}^{d-1}k_i$ with $k_i \in \{0, 1, \ldots, m\}$,
which in turn is the number of possible $m$-qudit initial states $|\re(\vec k)\rangle$; and, since $W_m^{(d)}$ is defined \eqref{Wop} by its action on $|\re(\vec k)\rangle$,
one can indeed naively expect to implement $W_m^{(d)}$ by using one $V$ operator for each possible state $|\re(\vec k)\rangle$.

As a further check, let us verify that we can recover our previous results for the qubit and qutrit cases.
For the case $d=2$, Eqs. \eqref{Wmd} and \eqref{Wmdj} reduce to
\begin{equation}
W_m^{(2)} = W_m^{(2,2)} =  
\overset{\curvearrowleft}{\prod_{l=1}^{m-1}} V_{m,l}^{(0,1)}\,,
\end{equation}
with $V_{m,l}^{(0,1)} = \RN{1}_{m,l}$ in Fig. \ref{fig:Iops}, which coincides with \eqref{W2explicit}.
For the case $d=3$, Eqs. \eqref{Wmd} and \eqref{Wmdj} reduce to
\begin{equation}
    W_m^{(3)} = W_m^{(3,3)}\, W_m^{(3,2)} \,,
\end{equation}
and 
\begin{align}
    W_m^{(3,2)} &= \overset{\curvearrowleft}{\prod_{l=1}^{m-1}}
\left(V_{m,l}^{(1,2)}\, V_{m,l}^{(0,2)}\, V_{m,l}^{(0,1)} \right) \,, \non\\
    W_m^{(3,3)} &= \overset{\curvearrowleft}
{\prod_{l_2=1}^{m-2}\prod_{l_1=l_2+1}^{m-1}} V_{m,l_1,l_2}^{(0,1,2)}
 \,,
\end{align}
with $V_{m,l_1,l_2}^{(0,1,2)}=\RN{2}_{m,l_1, l_2}$, which coincide with Eqs. 
\eqref{W3final}, \eqref{W3special}, \eqref{W3generic}, respectively, since 
$ W_m^{(3,2)}=\widetilde{W}_m$ and $W_m^{(3,3)}=\widetilde{\widetilde{W}}_{m}$.

We observe that the number of $V$ operators in $U_n^{(d)}$ is given by
\begin{equation}
\sum_{m=2}^n {m+d-1 \choose d-1} = \frac{n+1}{d}{n+d\choose d-1} -d-1 
= {\cal O}(n^d) \,,
\label{size}
\end{equation}
see Eqs. \eqref{Uresultd}, \eqref{weakcomps}. Each $V^{(i_0, i_1, \dots, i_{j-1})}_{m, l_1, \ldots, l_{j-1}}$ operator consists of $2(j-1)$ CNOT gates, and $j-1$ $(2j-1)$-fold controlled $R^y$ gates, as can be seen from Fig. \ref{fig:dgen}; the total number of such gates in $U_n^{(d)}$ is also ${\cal O}(n^d)$, as is the circuit depth. As previously noted, multi-controlled qudit $R^y$ gates can be decomposed into elementary 1-qudit and 2-qudit gates in the same way as for corresponding  multi-controlled qubit ($d=2$) $R^y$ gates.

The operator $U_n^{(d)}$ generates the qudit Dicke state $|D^n(\vec k)\rangle$ for any 
$\vec k$, see \eqref{DickeOpd}. For a fixed $\vec k$, we expect that it should be possible to prune away redundant $V$ operators and lower the circuit size, perhaps to ${\cal O}(l^{d-1} n)$ for $l_1 \sim l_2 \sim \ldots \sim l_{d-1} \equiv l \ll n$,
as we have done for the cases $d=2$ and $d=3$. However, we shall not pursue here such simplification for general values of $d$.

\section{Discussion}\label{sec:discuss}

We have formulated an algorithm for preparing qudit Dicke states. Our main results are the expression \eqref{Uresult} for the qudit Dicke operator $U_n$
as a product of $W$ operators \eqref{Wop}, and the decomposition of the $W$ 
operators in terms of elementary gates \eqref{W2explicit}, \eqref{W3final}, \eqref{Wmd}. 
For the qubit and qutrit cases, we have found simplified versions of these circuits, see
\eqref{Dicke2simple}-\eqref{W2explicit2} and 
\eqref{Dicke3simple}-\eqref{kkdefs};
and we have implemented these circuits in cirq.
The algorithm is deterministic, and does not use ancillary qudits.

Whereas the preparation of arbitrary qubit Dicke states has been considered in a 
number of works (see \cite{ 
Bartschi2019, Mukherjee:2020, Aktar:2021, Bartschi:2022} and references therein), ours is the 
first work (to our knowledge) to consider the preparation of 
arbitrary qudit Dicke states. We have seen that, already for the 
qutrit case, the algorithm entails a nontrivial generalization of 
\cite{Bartschi2019}. For the explicit gate implementation of the $W$ 
operators, we have aimed primarily for clarity rather 
than economy; it is likely that alternative implementations with 
reduced gate counts can be found. 

Having in hand a way to prepare qudit Dicke states,
one can begin to investigate their potential applications, 
such as those noted in the Introduction. In particular, 
it would be interesting to formulate an algorithm 
for preparing eigenstates of higher-rank integrable spin chains based on 
coordinate Bethe ansatz \cite{Sutherland:1975vr,Sutherland:1985}, thereby extending the 
approach \cite{VanDyke:2021kvq} for preparing eigenstates of
the Heisenberg spin chain.

Significant progress has recently been achieved on building quantum 
computers based on qutrits and even higher-dimensional qudits, see 
e.g. \cite{Goss:2022, Hrmo:2023, Morvan:2021, Ringbauer:2022, Roy:2022} and 
references therein. Since high-fidelity single-qutrit operations are 
already available \cite{Morvan:2021}, the initial (separable) state 
\eqref{qutritinitstate} can presumably already be implemented for a 
small number of qutrits. The generation of the simplest
generic qutrit Dicke state $|D^{3}(1,1,1)\rangle$ \eqref{D111} requires
up to double-controlled-rotation gates. Since 
high-fidelity single-controlled gates are already available \cite{Goss:2022, 
Hrmo:2023, Ringbauer:2022, Roy:2022}, it may be possible to implement 
this Dicke state in the near future.

\section*{Acknowledgments} 
We thank Hai-Rui Wei, Huangjun Zhu and Sreetama Das for helpful correspondence. This research was supported in part by the National Science Foundation under grants NSF PHY-1748958 and NSF 2310594, and by a Cooper fellowship.

\appendix

\section{Matrix representations of qutrit gates}\label{sec:matrices}

A 1-qutrit state lives in the 3-dimensional complex vector space $V$ spanned by ${|0\rangle, 
|1\rangle, |2\rangle}$. Let us set 
\begin{equation}
	|0\rangle = \begin{pmatrix}1\\0\\0\end{pmatrix} \,, \qquad
	|1\rangle = \begin{pmatrix}0\\1\\0\end{pmatrix} \,, \qquad 
	|2\rangle = \begin{pmatrix}0\\0\\1\end{pmatrix} \,.
\end{equation}
The NOT gates $X^{(ij)}$ \eqref{NOTgates} are represented by the $3 
\times 3$ matrices \cite{Di:2011}
\begin{equation}
X^{(01)}=\begin{pmatrix}
0 & 1 & 0\\
1 & 0 & 0\\
0 & 0 & 1
\end{pmatrix} \,, \qquad
X^{(02)}=\begin{pmatrix}
0 & 0 & 1\\
0 & 1 & 0\\
1 & 0 & 0
\end{pmatrix} \,, \qquad
X^{(12)}=\begin{pmatrix}
1 & 0 & 0\\
0 & 0 & 1\\
0 & 1 & 0
\end{pmatrix} \,.
\end{equation}
Similarly, the rotation gates $R^{(ij)}(\theta)$ \eqref{Rgates} are represented by
\begin{align}
R^{(01)}(\theta) &=\begin{pmatrix}
\cos(\frac{\theta}{2}) & -\sin(\frac{\theta}{2}) & 0\\[0.1 cm]
\sin(\frac{\theta}{2}) & \cos(\frac{\theta}{2}) & 0\\
0 & 0 & 1
\end{pmatrix} \,, \qquad
R^{(02)}(\theta)=\begin{pmatrix}
\cos(\frac{\theta}{2}) & 0 & -\sin(\frac{\theta}{2})\\
0 & 1 & 0\\
\sin(\frac{\theta}{2}) & 0 & \cos(\frac{\theta}{2})
\end{pmatrix} \,, \non \\
&\qquad\qquad R^{(12)}(\theta) =\begin{pmatrix}
1 & 0 & 0\\
0 & \cos(\frac{\theta}{2}) & -\sin(\frac{\theta}{2})\\[0.1 cm]
0 & \sin(\frac{\theta}{2}) & \cos(\frac{\theta}{2})
\end{pmatrix} \,.
\label{Rij}
\end{align}

An $n$-qutrit state lives in $V^{\otimes n}$. We label these vector 
spaces from 0 to $n-1$, going from {\em right to left}
\begin{equation}
\stackrel{\stackrel{n-1}{\downarrow}}{V} \otimes
\cdots 
\otimes
\stackrel{\stackrel{1}{\downarrow}}{V} \otimes
\stackrel{\stackrel{0}{\downarrow}}{V} \,. 
\end{equation}
In circuit diagrams, the $n$ vector spaces are represented by $n$ horizontal 
wires, which are labeled from 0 to $n-1$, going from top (0) to 
bottom $(n-1)$.
We use subscripts to indicate the vector spaces on which operators 
act nontrivially. For example, if $A$ is a 1-qutrit operator, then
\begin{equation}
A_{q} = \id^{\otimes (n-1-q)} \otimes A \otimes \id^{\otimes q} 
\end{equation}
is an operator on $V^{\otimes n}$ acting nontrivially on the $q^{th}$ vector space 
$q \in \{0, 1, \ldots, n-1\}$. The vector space on which an operator 
acts nontrivially can be changed 
using the permutation operator ${\cal P}_{q q'}$, for example
\begin{equation}
A_{q'} =  {\cal P}_{q\, q'}\, A_{q}\, {\cal P}_{q\, q'}\,,
\end{equation}
where
\begin{equation}
{\cal P}_{q\, q'} = \sum_{i, j = 1}^{3} e^{i,j}_{q}\, e^{j,i}_{q'} \,,
\end{equation}
and $e^{i,j}$ is the elementary $3 \times 3$ matrix whose $(i,j)$ matrix 
element is 1, and all others are 0; that is, $(e^{i,j})_{a,b} = 
\delta_{i,a}\, \delta_{j,b}$.

\begin{figure}[htb]
	\centering
	\begin{subfigure}{0.5\textwidth}
      \centering
\begin{adjustbox}{width=1.0\textwidth}
\begin{quantikz}
\lstick{$0$} & \gate{X^{(i,j)}} \vqw{1} & \qw \\
\lstick{$1$} & \gateO{1} & \qw 
\end{quantikz}
=
\begin{quantikz}
\lstick{$0$} &\qw & \gate{X^{(i,j)}} \vqw{1} & \qw & \qw \\
\lstick{$1$} &\gate{X^{(1,2)}} & \gateO{2} &\gate{X^{(1,2)}} & \qw 
\end{quantikz}
\end{adjustbox}
	  \caption{Identity for $C^{[1]}_{1}X^{(ij)}_{0}$}
	  \label{fig:id12}
    \end{subfigure}%
    \begin{subfigure}{0.5\textwidth}
        \centering
\begin{adjustbox}{width=1.0\textwidth}
\begin{quantikz}
\lstick{$0$} & \gate{X^{(i,j)}} \vqw{1} & \qw \\
\lstick{$1$} & \gateO{0} & \qw 
\end{quantikz}
=
\begin{quantikz}
\lstick{$0$} &\qw & \gate{X^{(i,j)}} \vqw{1} & \qw & \qw \\
\lstick{$1$} &\gate{X^{(0,2)}} & \gateO{2} &\gate{X^{(0,2)}} & \qw 
\end{quantikz}
\end{adjustbox}
	    \caption{Identity for $C^{[0]}_{1}X^{(ij)}_{0}$}
        \label{fig:id02}
	 \end{subfigure}	
\caption{Circuit diagrams for identities \eqref{CNOTident}}	 
\end{figure}
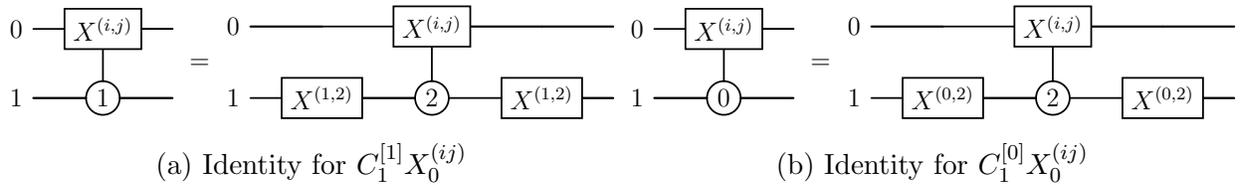

For the controlled-$X^{(ij)}$ gates \eqref{CX}, we have the $9 \times 
9$ block-diagonal matrices
\begin{equation}
C^{[2]}_{1}X^{(ij)}_{0} = \begin{pmatrix}
\myone^{6} &  \\
        & X^{(ij)}
\end{pmatrix} \,, \qquad 
C^{[1]}_{1}X^{(ij)}_{0} = \begin{pmatrix}
\myone^{3} &          &  \\
        & X^{(ij)} &  \\
        &          & \myone^{3}
\end{pmatrix} \,, \qquad 
C^{[0]}_{1}X^{(ij)}_{0} = \begin{pmatrix}
X^{(ij)} &  \\
         & \myone^{6}
\end{pmatrix} \,,
\end{equation}
where $\myone^{n}$ denotes the $n \times n$ identity matrix. The 
controlled gates are related by NOT gates on the controls, for example
\begin{equation}
C^{[1]}_{1}X^{(ij)}_{0} =  	X^{(12)}_{1}\, 
\left(C^{[2]}_{1}X^{(ij)}_{0}\right)\, 
X^{(12)}_{1} \,, \qquad
C^{[0]}_{1}X^{(ij)}_{0} =  	X^{(02)}_{1}\, 
\left(C^{[2]}_{1}X^{(ij)}_{0}\right)\, 
X^{(02)}_{1} \,.
\label{CNOTident}
\end{equation}
In terms of circuit diagrams, these identities are shown in Figs. 
\ref{fig:id12} and \ref{fig:id02}, respectively.

\noindent
Double-controlled-$X^{(ij)}$ gates are given by $3^{3} \times 
3^{3}$ block-diagonal matrices
\begin{equation}
C^{[22]}_{21}X^{(ij)}_{0} = \begin{pmatrix}
\myone^{24} &  \\
        & X^{(ij)}
\end{pmatrix} \,, \qquad 
C^{[11]}_{21}X^{(ij)}_{0} = \begin{pmatrix}
\myone^{12} &          &  \\
        & X^{(ij)} &  \\
        &          & \myone^{12}
\end{pmatrix} \,, \qquad 
C^{[00]}_{21}X^{(ij)}_{0} = \begin{pmatrix}
X^{(ij)} &  \\
         & \myone^{24}
\end{pmatrix} \,,
\end{equation}
and similarly for higher-controlled-$X^{(ij)}$ gates. 
Matrices corresponding to controlled $R^{y}$ rotation gates are defined in a similar 
way, with $X^{(ij)}$ replaced by $R^{(ij)}(\theta)$.


\providecommand{\href}[2]{#2}\begingroup\raggedright\endgroup

\end{document}